\documentclass{SCGE}
\usepackage{hyperref}
\begin{document}

%%%%%%新版式要加上这组
\begin{picture}(0,0){\rm
\put(0,-20){\makebox[160truemm][l]{\bf {\sanhao\raisebox{2pt}{.}}
Article  {\sanhao\raisebox{1.5pt}{.}}}}}
\put(0,-34){\jiuwuhao {\textcolor[rgb]{0.5,0.5,0.5}{\sf %Special Topic: Fluid Mechanics
}}}%%(11月注释：调\textcolor[rgb]{x,x,x}中的数字x越大越灰)
\end{picture}

\def\bm{\boldsymbol}

\def\dl{\displaystyle}
\def\du{\end{document}}
\def\d{{\rm d}}
\def\e{{\rm e}}
\def\i{{\rm i}}

% The author doesn't need fill in it.
\Year{} %
\Month{} %
\Vol{} %  卷号
\No{} %  期号
\BeginPage{1} % 起页码
\EndPage{??} %  止页码
\AuthorMark{{\rm Gao M}, et al.}  %(11月注释：页眉上的作者)
\AuthorMarkCite{{\rm Gao M}, et al.} %(11月注释：citation中的作者)
\DOI{} % The author doesn't need fill in it.

\title{Dynamics and entanglement of a membrane-in-the-middle optomechanical system in the extremely-large-amplitude regime}%标题

\author[1]{Gao Ming}{}
\author[1]{Lei FuChuan}{}
\author[1]{Du ChunGuang}{}
\author[1,2,3]{Long GuiLu}{Corresponding author (email: gllong@mail.tsinghua.edu.cn)}

\address[{\rm1}]{State Key Laboratory of Low-dimensional Quantum Physics and Department of Physics, Tsinghua University, Beijing 100084, China;}
\address[{\rm2}]{Tsinghua National Laboratory of Information Science and Technology, Beijing 100084, China;}
\address[{\rm3}]{Collaborative Innovation Center of Quantum Matter, Beijing 100084, China}

\maketitle \vspace{-3.5mm}{\footnotesize\begin{center} Received Month date, Year; accepted Month date, Year%收稿日期
\end{center}}\vspace*{-5mm}

%     Abstract is required.
\begin{center}
\rule{16.5cm}{0.4pt}
\parbox{16.5cm}
{\begin{abstract}
The study of optomechanical systems has attracted much attention, most of which are concentrated in the physics in the small-amplitude regime.
While in this article, we focus on optomechanics in the extremely-large-amplitude regime and consider both classical and quantum dynamics.
Firstly, we study classical dynamics in a membrane-in-the-middle optomechanical system in which a partially reflecting and flexible membrane is suspended inside an optical cavity.
We show that the membrane can present self-sustained oscillations with limit cycles in the shape of sawtooth-edged ellipses and exhibit dynamical multistability.
Then, we study the dynamics of the quantum fluctuations around the classical orbits.
By using the logarithmic negativity, we calculate the evolution of the quantum entanglement between the optical cavity mode and the membrane during the mechanical oscillation.
We show that there is some synchronism between the classical dynamical process and the evolution of the quantum entanglement.
\end{abstract}}
\end{center}\vspace*{-0.6cm}

\begin{center}
\parbox{16.5cm}
{\bf\jiuhao optomechanics, self-sustained oscillation, entanglement, membrane-in-the-middle optomechanical system, MIMOS, extremely-large-amplitude regime, ELAR}%关键词
\end{center}

\begin{center}
{}%分类号
%\CITA    %%(11月注释：Citation内容自动生成)
\Cit{Gao M, Lei F C, Du C G, Long G L. Dynamics and entanglement of a membrane-in-the-middle optomechanical system in the extremely-large-amplitude regime. Sci China-Phys Mech Astron, 2016 , : , DOI: 10.1007/s11433-015-5704-5}%%(11月注释：Citation内容需手动填写)
\end{center}

\textwidth=178truemm \textheight=236truemm%%%%%%新版式要加上

%%%%%%%%%%%%%%%%%%%%%%%%%%%%%%%%%%%%%%%%%%%%%%%%%%%%%%%%%%%%
\wuhao\vspace*{1.5mm}

\begin{multicols}{2}

%%%%%%%%%%%%%%%%%%%%%%%%%%%%%%%%%%%%%%%%%%%%%%%%%%%%%%%%%%%%
  %% Text of article.
%%%%%%%%%%%%%%%%%%%%%%%%%%%%%%%%%%%%%%%%%%%%%%%%%%%%%%%%%%%%
  % Section headings
  \renewcommand{\baselinestretch}{1.08} \baselineskip
  12.2pt\parindent=10.8pt

  \renewcommand{\thefootnote}

%%%%%%%%%%%%%%%%%%%%%%%%%%%%%%%%%%%%%%%%%%%%%%%%%%%%%%%%%%%%%%%%%%%%%%%%%%%%%%%%%%%%%%%%
\section{INTRODUCTION}

Optomechanics \cite{review1,review2} concerns with the coupling between optical cavity modes  and mechanical degrees of freedom via radiation pressure \cite{radiation-prl-1999}, optical gradient forces \cite{gradient-force}, photothermal forces \cite{photothermal-nature-2004} or the Doppler effect \cite{doppler-prl-2008}.
Optomechanical systems have potential applications in many areas, such as detection of gravitational wave \cite{gravitational-pla-2001,gravitational-pla-2002,gravitational-pla-2002-2}, sensitive sensors \cite{sensor-prl-2006,lsawater,scpmaresponse}, mechanical memory \cite{memory-nnano-2011}, cooling of mechanical resonators \cite{cooling-mechanical,cooling-mechanical2}, coherent manipulation of light \cite{transparency1,wuli1}, studies of quantum entanglement \cite{testquantum-prl-2007,entanglements1,entanglements2,entanglements3}, preparation of macroscopic quantum state \cite{quantum-state-prl-2002,aplhwy3wave,praliuyx} and quantum information processing \cite{information-prl-2010,lei-engineering}.

Most early works focus attention on optomechanics in the small-amplitude regime, in which the optomechanical coupling can be approximately considered as linear or quadratic.
With a driving laser of frequency blue detuned with respect to the cavity resonance and power above a certain threshold, the mechanical resonator can run into self-sustained oscillations \cite{chaos-prl-2005,sso-prl-2005,sso-oe-2005} and dynamical multistability may emerge \cite{multistability-prl-2006,multistability-prl-2008}.
The mechanical resonator conducts approximately sinusoidal oscillations, and therefore, its limit cycles in the phase space are approximately elliptical.

In the large-amplitude regime, which can be realized in an optomechanical system driven by a high-power laser, the mechanical resonator can display different self-sustained oscillations with limit cycles being mushroom-like in shape \cite{large-amplitude-pra-2012}.
If the power of the driving laser is further increased, the amplitude of the mechanical oscillation can be comparable with the wavelength of the laser, i.e., the system reaches the extremely-large-amplitude regime (ELAR) \cite{ELAR}.
In this regime, the optomechanical coupling should not be simply considered as linear or quadratic, but should be treated directly as a exact function of the position of the mechanical resonator without approximation.
In our previous work \cite{ELAR}, we studied purely classical dynamics in the ELAR of an optomechanical system which is a Fabry-P\'{e}rot cavity with a perfectly reflecting and movable end mirror on one side.
In that model, multiple optical cavity modes of different orders may be excited and coupled with the movable mirror via radiation pressure during the mechanical oscillation in the ELAR.

In this article, we consider another optomechanical model in which a partially reflecting and flexible membrane is placed inside a Fabry-P\'{e}rot cavity.
Recently, this type of membrane-in-the-middle optomechanical system (MIMOS) has been widely investigated \cite{membrane-nature-2008,membrane-np-2010}, due to their superiority to the traditional counterparts.
The optical and mechanical functionality are segregated to physically distinct structures, so both excellent optical and mechanical properties can be achieved simultaneously.
In addition, when the membrane is placed at a node (or anti-node) of the intracavity standing wave, the system can exhibit quadratic optomechanical coupling \cite{membrane-nature-2008}, this allows for direct measurement of the square of the membrane's displacement, and thus quantum non-demolition readout of the membrane's eigenstates.
In this opomechanical model, the cavity resonant frequencies are periodic in the membrane's position.
In the ELAR, with an external driving laser of fixed frequency, one single cavity mode is excited multiple times (rather than multiple cavity modes are excited two times as in \cite{ELAR}) and coupled with the membrane during one cycle of the mechanical oscillation.
In this article, we study classical dynamics of the system  and the evolution of the quantum fluctuations around the classical orbits in the ELAR. We also calculate the entanglement between the optical cavity mode and the membrane in the ELAR, as entanglement is an important and precious resource for quantum information processing tasks\cite{tele,dense,qsdc-1,qsdc-2,qsdc-3,computation1,computation2,informationadd}.

We organize the article in the following way.
Sec. \ref{sec2} introduces the model and the Hamiltonian, and gives the dynamical equations of the system.
In Sec. \ref{sec3}, the classical dynamics of the system is studied by analyzing the limit cycles of the mechanical oscillation of the membrane in the phase space.
The evolution of the quantum fluctuations around the classical orbits and the quantum entanglement between the optical cavity mode and the mechanical resonator is studied in Sec. \ref{sec4}.
Finally, Sec. \ref{sec5} gives a summary of this work.

 %%%%%%%%%%%%%%%%%%%%%%%%%%%%%%%%%%%%%%%%%%%%%%%%%%%%%%%%%%%%%%%%%%%%%%%%%%%%%%%%%%%%%%%%%%%%%%%%%%%%%%%
\section{Hamiltonian and Dynamical Equations}
\label{sec2}

\begin{figure}[H]
\centering
\includegraphics[width=3.1 in]{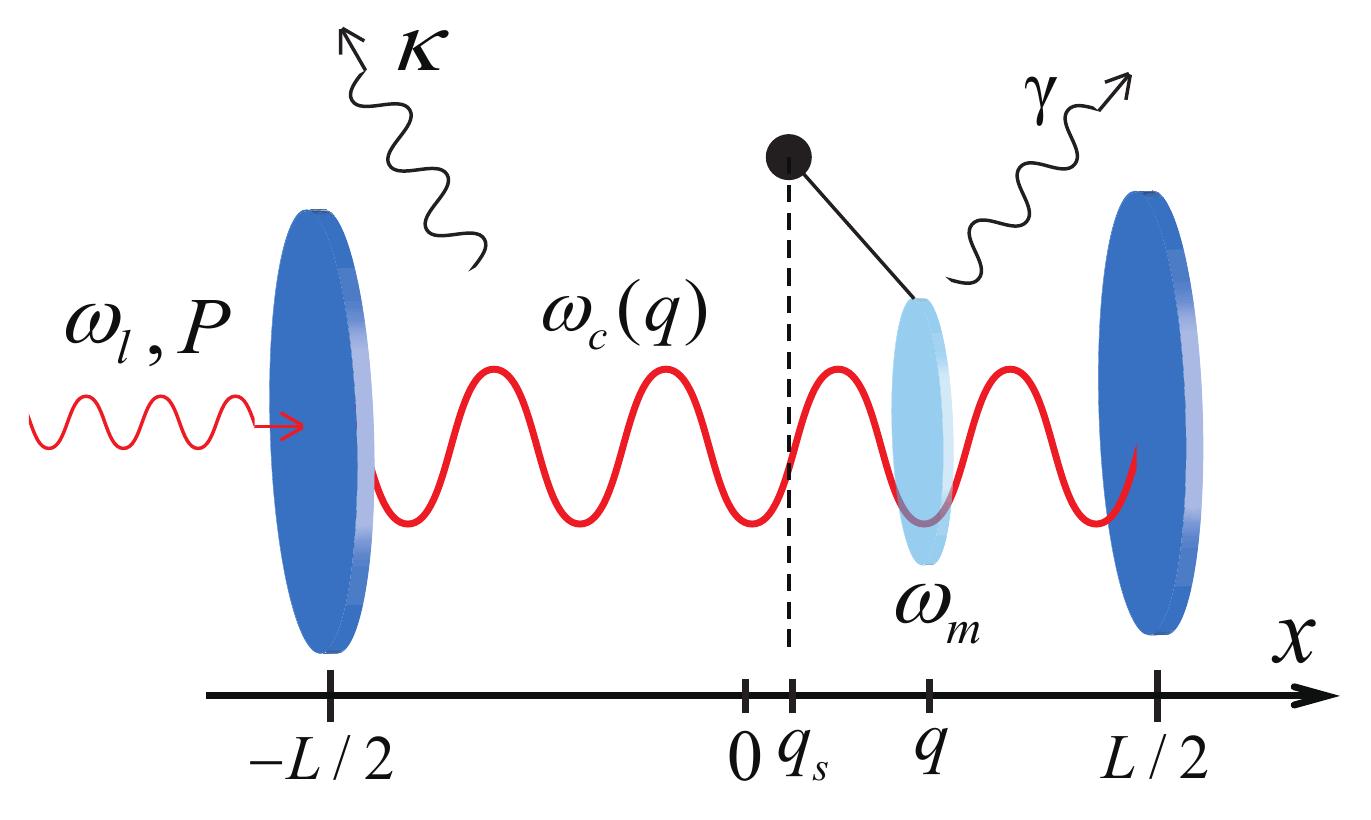}
\caption{(Color online) Schematic of a MIMOS. A partially reflecting, flexible thin dielectric membrane is placed inside a Fabry-P\'{e}rot cavity.
The cavity of length $L$ is driven by an external laser of frequency $\omega_l$ and power $P$.
The membrane can be considered as a mechanical resonator with intrinsic frequency $\omega_m$, damping rate $\gamma$, static equilibrium position $q_s$ and dynamic position $q$.
It is coupled with a cavity mode with $q$-dependent frequency $\omega_c(q)$ and decay rate $\kappa$ via radiation pressure.}
\label{fig1}
\end{figure}

We consider a MIMOS as shown in Fig. \ref{fig1}, in which a partially reflecting, flexible thin dielectric membrane is placed inside a Fabry-P\'{e}rot cavity.
The flexible membrane in this model can be considered as a mechanical resonator of intrinsic frequency $\omega_m$, mass $m$ and damping rate $\gamma$.
The cavity is of length $L$ with the end mirrors fixed at $x=\pm L/2$.
It is driven by an external laser of frequency $\omega_l$ and power $P$.
If the driving laser is turned off, the static equilibrium position of the membrane is $q_s$.

\begin{figure}[H]
\centering
\includegraphics[width=3.3 in]{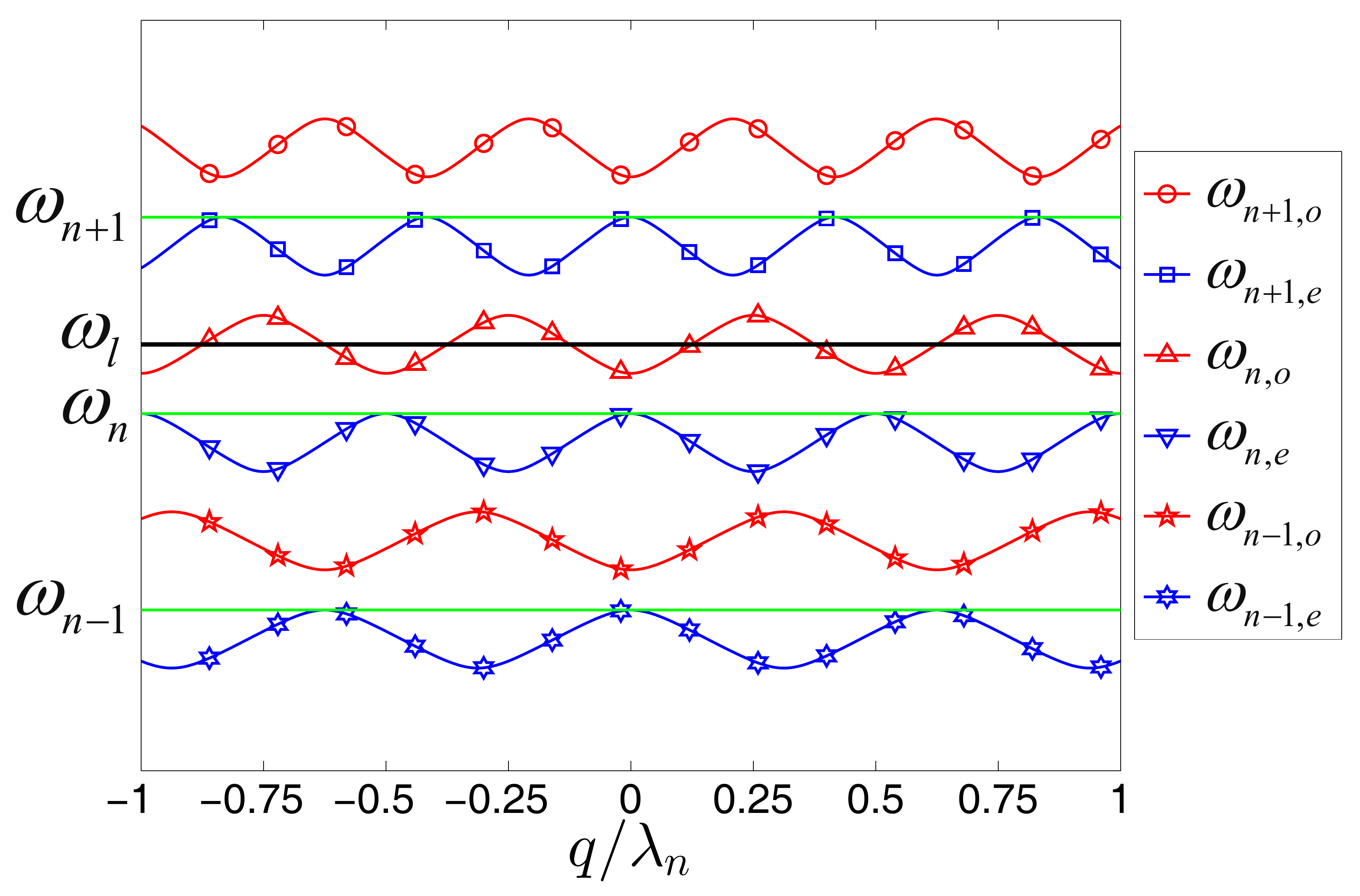}
\caption{(Color online) The frequencies of the cavity modes, which are periodic in the position of the membrane.}
\label{fig2}
\end{figure}

In this system, radiation pressure provides the dominant optomechanical coupling which is typically dispersive, implying that the primary effect of the membrane's motion is to shift the frequencies of the optical cavity modes.
In the simplest case that $|r_c|=1$ and $q=0$, where $r_c$ is the reflectivity of the membrane and $q$ is its position, the cavity is divided into two uncoupled subcavities with equal lengths $L/2$ by the membrane.
The resonant frequencies of the two subcavities are both $\omega_n=2n\pi c/L$ with mode number $n$ and wavelength $\lambda_n=L/n$.
In the general case, the membrane is partially reflecting, i.e., $|r_c|\not= 1$, the two subcavities are coupled and each pair of the twofold degenerate modes of frequencies $\omega_n$ splits into a pair of nondegenerate modes of frequencies $\omega_{n,e}$ and $\omega_{n,o}$ as below \cite{membrane-pra-2008},
\begin{align}
\label{eq1}
\omega_{n,e}(q)&=\omega_n+\frac{c}{L}\sin^{-1}\left[|r_c|\cos(\frac{4\pi q}{\lambda_n})\right]-\frac{c}{L}\sin^{-1}(|r_c|),\\
\label{eq2}
\omega_{n,o}(q)&=\omega_n+\frac{\pi c}{L}-\frac{c}{L}\sin^{-1}\left[|r_c|\cos(\frac{4\pi q}{\lambda_n})\right]-\frac{c}{L}\sin^{-1}(|r_c|).
\end{align}
Thus, the frequencies of the optical cavity modes are periodic functions of $q$, as shown in Fig. (\ref{fig2}).
If one cavity mode of frequency $\omega_c(q)=\omega_{n,\sigma}(\sigma=e,o)$ and decay rate $\kappa$ satisfies the below relations:
\begin{equation}
\label{ej1}
\min(\omega_{n,\sigma})\leq\omega_l\leq\max(\omega_{n,\sigma}),\nonumber
\end{equation}
and
\begin{equation}
\label{ej2}
\kappa\ll\frac{\pi c}{L}-\frac{2c}{L}\sin^{-1}(|r_c|),\nonumber
\end{equation}
where $\pi c/L-2c/L\sin^{-1}(|r_c|)$ is the minimum frequency difference between the cavity mode and the adjacent cavity modes,
then it is appropriate to consider only this single cavity mode participating the coupling with the membrane during the mechanical oscillation.

In most early works, the displacement $q-q_s$ of the membrane is assumed to be very small.
In this small-amplitude regime, $\omega_c(q)$ can be approximately expanded to at most the second order \cite{membrane-nature-2008} around the static equilibrium position $q_s$,
\begin{equation}
\omega_c(q)\approx\omega_{c0}+g_1(q-q_s)+g_2(q-q_s)^2,\nonumber
\end{equation}
where $g_1=\omega_c^{\prime}(q_s)$ is the linear coupling strength and $g_2=\omega_c^{\prime\prime}(q_s)/2$ is the quadratic coupling strength.

If the optical cavity is driven by a high-power laser, the small-amplitude assumption will be no longer valid and the expansion of $\omega_c(q)$ to the second-order will not be a good approximation.
It is necessary to deal with $\omega_c(q)$ directly without any approximate expansion.
If the power $P$ is high enough, the amplitude $A$ of the mechanical oscillation may be comparable with the wavelength $\lambda_l$ of the driving laser,
\begin{equation}
A/\lambda_l\sim 1,\nonumber
\end{equation}
i.e., the system reaches the ELAR.
In this case, as $\omega_c(q)$ is a periodic function of $q$, each time the membrane passes through the positions that satisfy $\omega_c(q)=\omega_l$, the optical cavity mode is excited.
So during one whole cycle of the mechanical oscillation, the optical cavity mode is excited multiple times and be coupled with the membrane via radiation pressure.
It should be noticed that, this case is different from the one in our previous work, Ref. \cite{ELAR}, in which the cavity resonant frequencies are monotonic functions of the position of the movable end mirror, so in the ELAR, multiple optical cavity modes of different orders are excited twice during one cycle of the mechanical oscillation.
Here, we assume that the size of the membrane is much larger than its amplitude of mechanical oscillation.
In this situation, we can treat the membrane as a harmonic resonator and neglect the effect of mechanical nonlinearities \cite{nonlinear-njp-2012}.
In the frame rotating at frequency $\omega_l$, the Hamiltonian of the system reads
\begin{align}
\label{eq3}
H=&\hbar[\omega_c(\hat{q})-\omega_l]\hat{a}^\dag \hat{a}+\frac{\hat{p}^2}{2m}+\frac{1}{2}m\omega_{m}^2 (\hat{q}-q_s)^2 \nonumber\\
&+\hbar\alpha_L(\hat{a}+\hat{a}^\dag)+H_{\kappa}+H_{\gamma} ,
\end{align}
where $\hat{a}$ and $\hat{a}^\dag$ are the bosonic annihilation and creation operators of the optical cavity mode, $\hat{q}$ and $\hat{p}$ are the position and momentum operators of the membrane, $\alpha_L$ is the complex amplitude of the driving laser field which satisfies $|\alpha_L|^2=2\kappa P/\hbar \omega_l$.
Here, without loss of generality, we set $\alpha_L$ to be real.
$H_{\kappa}$ denotes the coupling between the optical cavity mode and the vacuum bath that leads to the decay rate $\kappa$.
$H_{\gamma}$ refers to the interaction between the membrane and the thermal reservoir which is the cause of the damping rate $\gamma$.
The quantum Langevin equations (QLEs) for operators $\hat{q}$, $\hat{p}$ and $\hat{a}$ can be easily derived from the Hamiltonian as follows,
\begin{align}
\label{eq4}
\frac{d}{dt}\hat{q}&=\frac{\hat{p}}{m}, \\
\label{eq5}
\frac{d}{dt}\hat{p}&=-\hbar\omega_c^\prime(\hat{q})\hat{a}^\dag\hat{a}-m\omega_m^2(\hat{q}-q_s)-\gamma\hat{p}+\hat{\eta}, \\
\label{eq6}
\frac{d}{dt}\hat{a}&=-i[\omega_c(\hat{q})-\omega_l]\hat{a}-i\alpha_L-\kappa\hat{a}+\sqrt{2\kappa}\hat{a}_{in},
\end{align}
where $\hat{\eta}$ is the mechanical Brown noise operator and $\hat{a}_{in}$ is the optical vacuum input noise operator.
In the presence of strong external driving, we can rewrite each Heisenberg operator as the sum of classical mean value and quantum fluctuation operator as below,
\begin{align}
\hat{q}&=q_0(t)+q_{z}\delta\hat{q},\nonumber\\
\hat{p}&=p_0(t)+p_{z}\delta\hat{p},\nonumber\\
\hat{a}&=\alpha(t)+\delta\hat{a},\nonumber
\end{align}
where $q_{z}=\sqrt{\hbar/m\omega_m}$ and $p_{z}=\sqrt{\hbar m\omega_m}$ are the zero-point fluctuations of the membrane's position and momentum, respectively.
$\omega_c(\hat{q})$ and $\omega_c^\prime(\hat{q})$ in Eqs. (\ref{eq5})-(\ref{eq6}) can be approximately expanded around the classical mean position $q_0(t)$ as below,
\begin{align}
\omega_c(\hat{q})&=\omega_c(q_0)+\omega_c^\prime(q_0)q_{z}\delta\hat{q},\nonumber\\
\omega_c^\prime(\hat{q})&=\omega_c^\prime(q_0)+\omega_c^{\prime\prime}(q_0)q_{z}\delta\hat{q},\nonumber
\end{align}
Thus, we can obtain the classical dynamical equations of the system:
\begin{align}
\label{eq7}
\frac{d}{dt}q_0&=\frac{p_0}{m}, \\
\label{eq8}
\frac{d}{dt}p_0&=-\hbar\omega_c^\prime(q_0)|\alpha|^2-m\omega_m^2(q_0-q_s)-\gamma p_0,\\
\label{eq9}
\frac{d}{dt}\alpha&=-i[\omega_c(q_0)-\omega_l]\alpha-i\alpha_L-\kappa\alpha,
\end{align}
and the QLEs for the quantum fluctuations:
\begin{align}
\label{eq10}
\frac{d}{dt}\delta\hat{q}=&\omega_m\delta\hat{p}, \\
\label{eq11}
\frac{d}{dt}\delta\hat{p}=&-\omega_m\delta\hat{q}-q_{z}^2\omega_c^{\prime\prime}(q_0)|\alpha|^2\delta\hat{q}-q_{z}\omega_c^\prime(q_0)\alpha^*\delta\hat{a}\nonumber\\&-q_{z}\omega_c^\prime(q_0)\alpha\delta\hat{a}^\dag-\gamma\delta\hat{p}+\hat{\xi}, \\
\label{eq12}
\frac{d}{dt}\delta\hat{a}=&-i[\omega_c(q_0)-\omega_l]\delta\hat{a}-iq_{z}\omega_c^\prime(q_0)\alpha\delta\hat{q}-\kappa\delta\hat{a}\nonumber\\&+\sqrt{2\kappa}\hat{a}_{in},
\end{align}
where $\hat{\xi}=\hat{\eta}/q_{z}$.
The noise operators $\hat{\xi}$ and $\hat{a}_{in}$ have zero mean values and are characterized by their auto correlation functions \cite{markovianp}
\begin{equation}
\label{eq13}
\left\langle\hat{a}_{in}(t)\hat{a}^\dag_{in}(t^\prime)\right\rangle =\delta(t-t^\prime),
\end{equation}
and
\begin{equation}
\label{eq14}
\left\langle\hat{\xi}(t)\hat{\xi}(t^\prime)\right\rangle =\frac{\gamma}{\omega_m}\int\frac{d\omega}{2\pi}e^{-i\omega(t-t^\prime)}\omega\left[\coth\left(\frac{\hbar\omega}{2k_{B}T}\right)+1\right],\nonumber\\
\end{equation}
where $k_B$ is the Blotzmann constant and $T$ is the temperature of the membrane.
The noise operator $\hat{\xi}$ is not delta-correlated and therefore does not describe a Markovian process \cite{markovianp}.
However, quantum effects are achievable only when the membrane has a high mechanical quality factor ($Q_m=\omega_m/\gamma\gg1$).
In this limit, $\hat{\xi}$ becomes delta-correlated \cite{noise-correlation},
\begin{equation}
\label{eq15}
\left\langle\hat{\xi}(t)\hat{\xi}(t^\prime)+\hat{\xi}(t^\prime)\hat{\xi}(t)\right\rangle /2\simeq\gamma(2n_{th}+1)\delta(t-t^\prime),
\end{equation}
where $n_{th}=[\exp(\hbar\omega_m/k_BT)-1]^{-1}$ is the mean thermal excitation number, and we recover a Markovian process.

%%%%%%%%%%%%%%%%%%%%%%%%%%%%%%%%%%%%%%%%%%%%%%%%%%%%%%%%%%%%%%%%%%%%%%%%%%%%%%%%%%%%%%
\section{classical dynamics}
\label{sec3}
\begin{figure}[H]
\centering
\includegraphics[width=3.4 in]{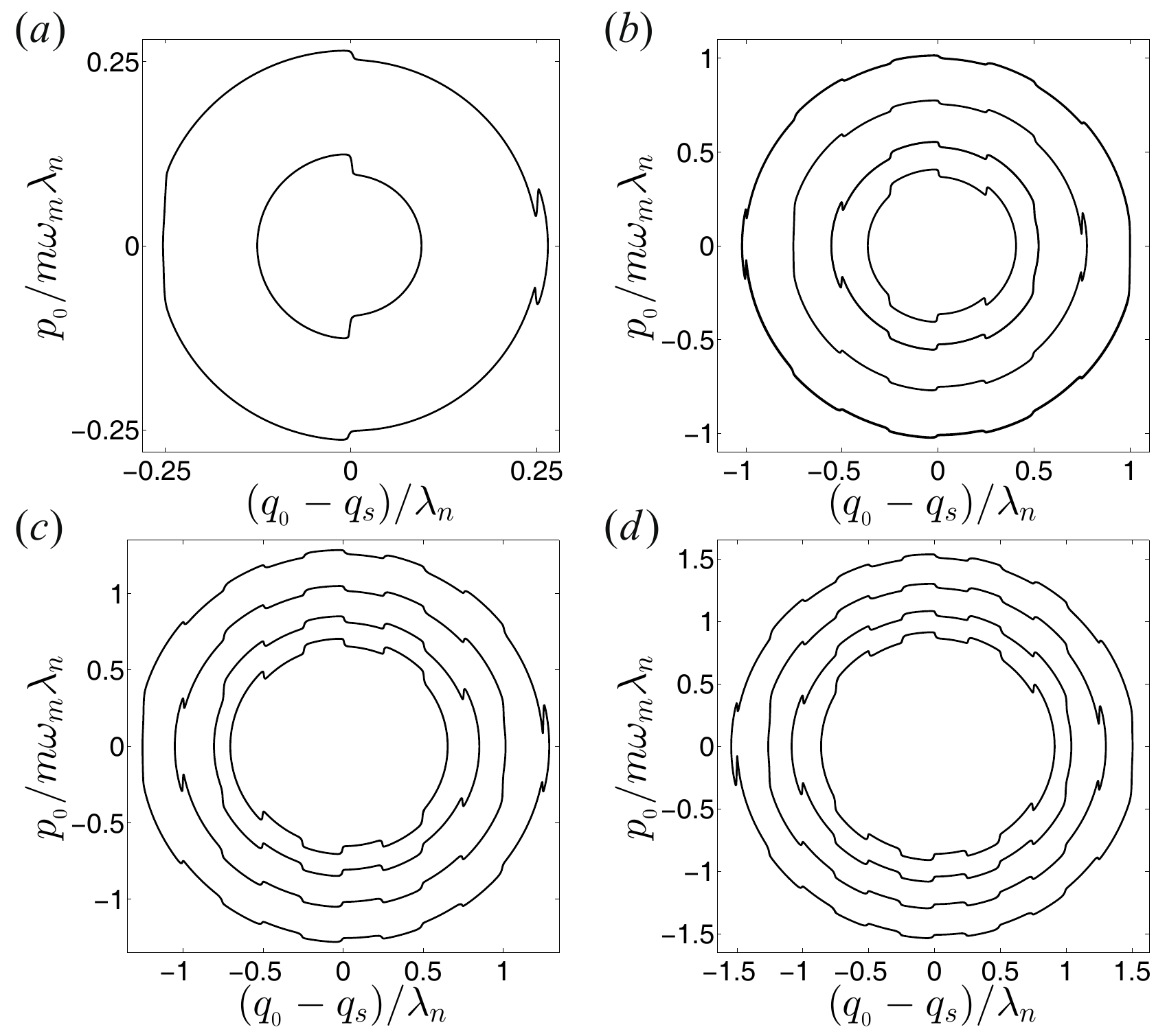}
\caption{Limit cycles in the phase space of the mechanical resonator scanned by $q_0$ and $p_0$ with different values of $P$, which are: (a) $P=0.02$ W, (b) $P=0.095$ W, (c) $P=0.21$ W, (d) $P=0.27$ W, as specified by the vertical black dashed lines in Fig. \ref{fig4}.
}
\label{fig3}
\end{figure}

Firstly, we focus on the purely classical dynamics of the system.
We integrate Eqs. (\ref{eq7})-(\ref{eq9}) numerically by using a fourth-order Runge-Kutta algorithm.
Considering the experimental feasibility \cite{membrane-nature-2008,membrane-np-2010}, we set the parameters as follows, the intrinsic frequency, mass, damping rate and reflectivity of the membrane are $\omega_m=2\pi\times10^5$ Hz, $m=5\times10^{-14}$ kg, $\gamma=10^{-2}\omega_m$ and $r_c=0.8$.
The length of the cavity is $L=6$ cm.
During the mechanical oscillation of the membrane, only one single cavity mode is excited, it is of order $n=60000$, wavelength $\lambda_n=1000$ nm, frequency $\omega_c=\omega_{n,o}$ and decay rate $\kappa=50\omega_m$.
It should be noticed that we set the cavity in the unresolved sideband regime ($\kappa>\omega_m$) to make sure it can respond quickly enough to the fast mechanical oscillation.
The static equilibrium position of the membrane is set to be $q_s=\lambda_n/8$.
The system is driven by a external laser of frequency $\omega_l=\omega_{n,o}(q_s)$.
\begin{figure}[H]
\centering
\includegraphics[width=3.3 in]{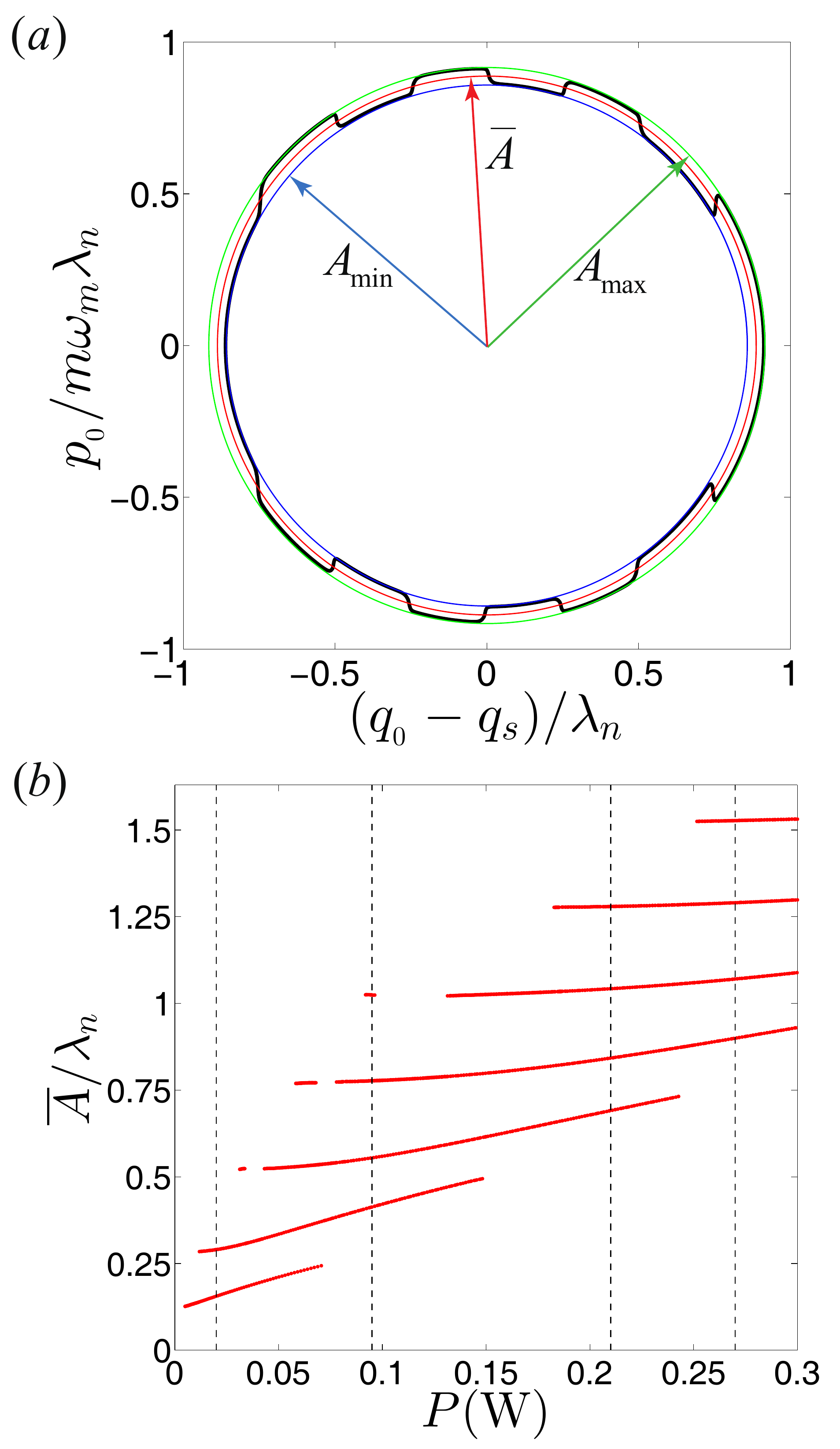}
\caption{(Color online) (a) Definition of the average amplitude as $\overline{A}=\sqrt{(A_{min}^2+A_{max}^2)/2}$.
(b) Attractor diagram on a plane of $\overline{A}$ and $P$.
The vertical black dashed lines specify some values of $P$ distributed in different regions.
Limit cycles for these values of $P$ are plotted in Fig. \ref{fig3}.
}
\label{fig4}
\end{figure}

From the numerical solutions of Eqs. (\ref{eq7})-(\ref{eq9}), we plot the limit cycles in the phase space of the mechanical resonator scanned by $q_0$ and $p_0$ as shown in Fig. \ref{fig3}.
Each limit cycle corresponds to a stable self-sustained oscillation in the long-time limit.
The limit cycles are in the shape of sawtooth-edged ellipses similar as the case in Ref. \cite{ELAR}.
However, there are some differences between the present limit cycles and the ones in Ref. \cite{ELAR}.
In Ref. \cite{ELAR}, the limit cycles expand when the movable mirror moves forward and shrink when the mirror moves backward.
It is because the cavity field is distributed on one side of the movable mirror.
When the mirror moves forward, the radiation pressure of the excited modes always does positive work on it and makes it sharply accelerate, as a reflection in the phase space, the limit cycles expand.
And vice versa, when the mirror moves backward, the limit cycles shrink.
While in the present model, the cavity field is distributed on both sides of the membrane.
When the membrane moves ,whether forward or backward, the radiation pressure of the excited cavity mode alternately does positive work and negative work on it.
So it experiences alternately sharp acceleration and deceleration.
As a reflection in the phase space, the limit cycles alternately expand and shrink whether along the forward direction or the backward direction.
\begin{figure*}[htp]
\centering
\includegraphics[width=5.1 in]{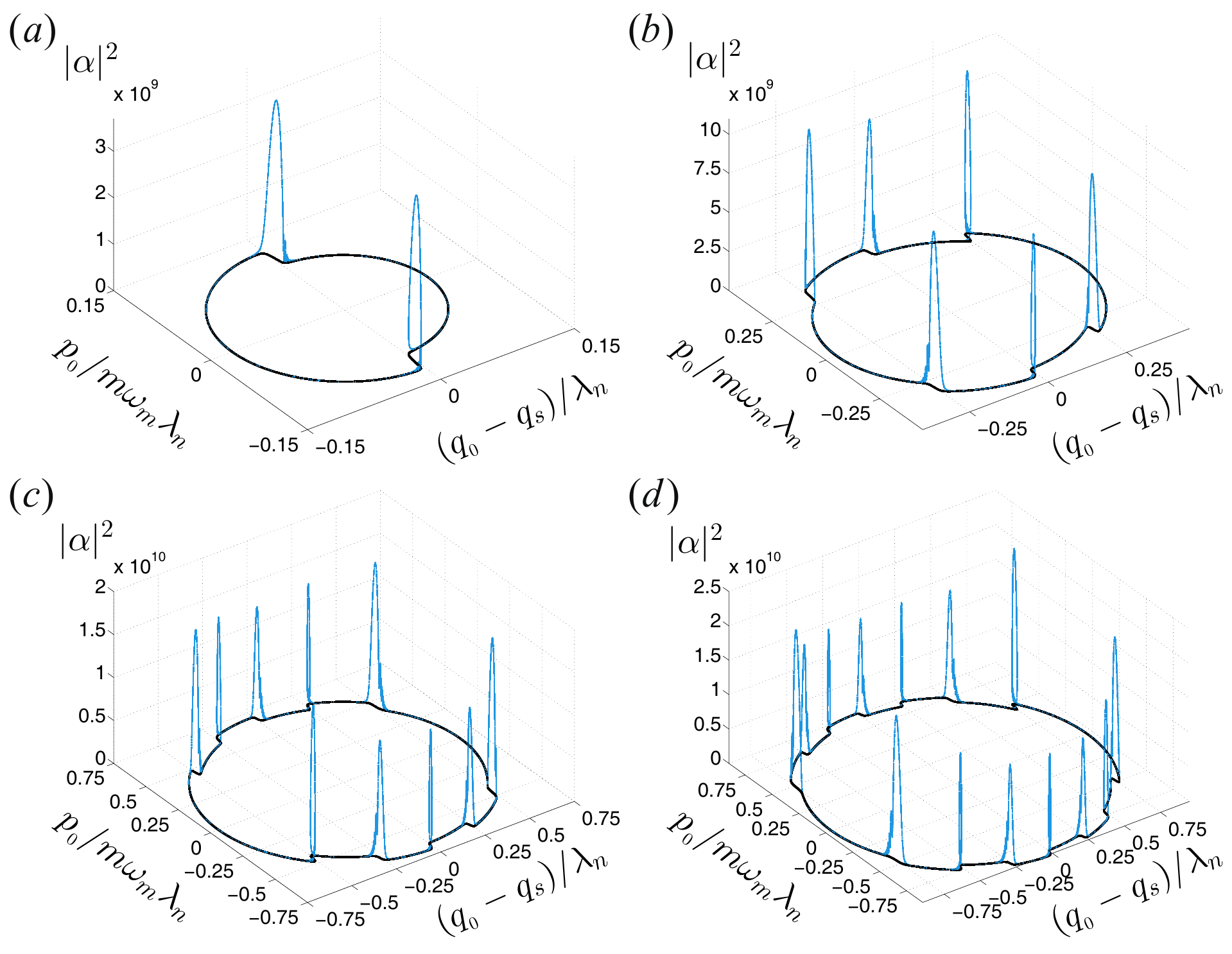}
\caption{(Color online) The variation of the optical cavity occupation with the mechanical oscillation of the membrane. The values of $P$ are: (a) $P=0.02$ W, (b) $P=0.095$ W, (c) $P=0.21$ W, (d) $P=0.27$ W, as specified by the vertical black dashed lines in Fig. \ref{fig4}. Here, we only consider the case that the membrane oscillating with the smallest limit cycles.
}
\label{fig5}
\end{figure*}

As shown in Fig. \ref{fig3}, there are multiple different limit cycles in the phase space, meaning that the mechanical resonator exhibits dynamical multistability.
The mechanical oscillation is stable only when the total net work done by radiation pressure is balanced with the dissipative energy during one whole cycle.
At some parameters, there are multiple stable oscillations that can satisfy the energy-balance condition.
So the mechanical resonator can exhibit dynamical multistability.

To demonstrate dynamical multistability concisely, we define average amplitude as $\overline{A}=\sqrt{(A_{min}^2+A_{max}^2)/2}$ as shown in Fig. \ref{fig4}(a), where $A_{min}$ and $A_{max}$ are respectively the minimum and maximum amplitude of a limit cycle.
It should be noticed that with the effect of radiation pressure, the dynamical equilibrium position $\overline{q_0}$ is shifted from the static equilibrium position $q_s$.
However, in the ELAR, this shift is very small compared with the amplitude and can be neglected.
So $A_{min}$, $A_{max}$, and $\overline{A}$ are all defined with the origin at the static equilibrium position $q_s$.
Fig. \ref{fig4}(b) shows the dependence of $\overline{A}$ on $P$, which can be considered as an attractor diagram.
The discreteness of $\overline{A}$ reveals that the energy-balance condition leads to an amplitude locking effect.
At some parameters, the mechanical resonator exhibits dynamical multistability, so $\overline{A}$ can take multiple values for a fixed $P$ as shown in Fig. \ref{fig4}(b).

We demonstrate the variation of the photon number in the cavity with the mechanical oscillation of the membrane in Fig. \ref{fig5}.
As the frequency of the cavity mode is periodic in the position of the membrane, every time the membrane passes though the positions that satisfy $\omega_c(q_0)=\omega_l$ (in our parametric space, that is $q_0=q_s+k\lambda_n/4$, $k\in\mathbb{N}$), the cavity mode is excited, and as a result, the photon number in the cavity experiences a maximum value.
\begin{figure*}[hbp]
\centering
\includegraphics[width=5.1 in]{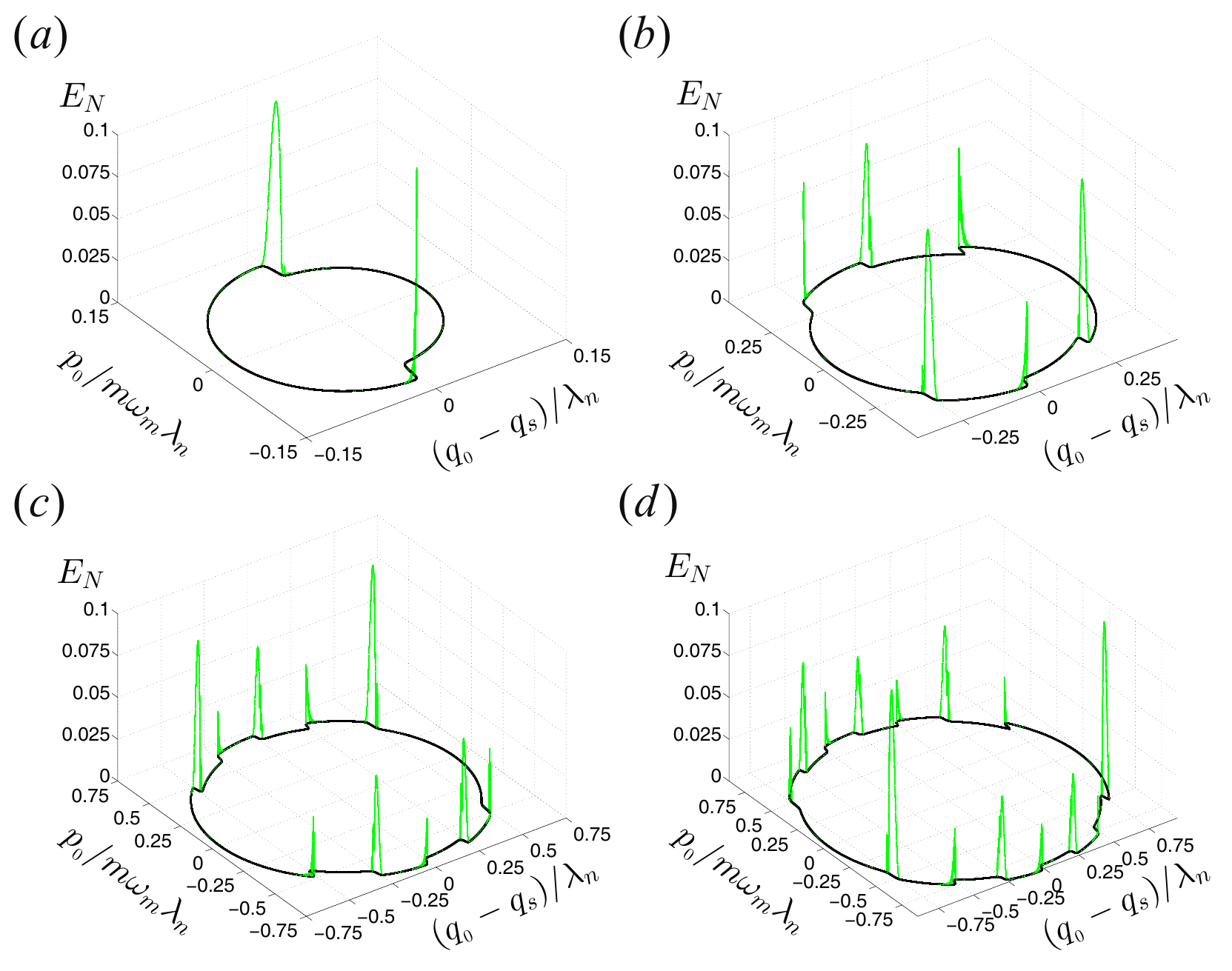}
\caption{(Color online) The variation of the quantum entanglement with the mechanical oscillation of the membrane. The values of $P$ are: (a) $P=0.02$ W, (b) $P=0.095$ W, (c) $P=0.21$ W, (d) $P=0.27$ W, as specified by the vertical black dashed lines in Fig. \ref{fig4}. Here, we only consider the case that the membrane oscillating with the smallest limit cycles. The temperature of the membrane is $T=1$ mK.
}
\label{fig6}
\end{figure*}

%%%%%%%%%%%%%%%%%%%%%%%%%%%%%%%%%%%%%%%%%%%%%%%%%%%%%%%%%%%%%%%%%%%%%%%%%%%%%%%%%%%%%%
\section{quantum entanglement}
\label{sec4}

Once the classical dynamics of the mean values $q_0(t)$, $p_0(t)$, and $\alpha(t)$ is obtained, we can solve the dynamics of the corresponding quantum fluctuations $\delta\hat{q}$, $\delta\hat{p}$, and $\delta\hat{a}$.
To proceed, we introduce the position and momentum quadratures for the cavity mode and the corresponding input noise:
\begin{align}
\delta\hat{x}&=\frac{\delta\hat{a}+\delta\hat{a}^\dag}{\sqrt{2}} ,\nonumber\\
\delta\hat{y}&=\frac{\delta\hat{a}-\delta\hat{a}^\dag}{i\sqrt{2}} ,\nonumber\\
\hat{x}_{in}&=\frac{\hat{a}_{in}+\hat{a}^\dag_{in}}{\sqrt{2}} ,\nonumber\\
\hat{y}_{in}&=\frac{\hat{a}_{in}-\hat{a}^\dag_{in}}{i\sqrt{2}} .\nonumber
\end{align}
We write all quadratures and noise operators in terms of vectors:
\begin{align}
u&=[\delta\hat{q},\delta\hat{p},\delta\hat{x},\delta\hat{y}]^T ,\nonumber\\
\epsilon &=[0,\hat{\xi},\sqrt{2\kappa}\hat{x}_{in},\sqrt{2\kappa}\hat{y}_{in}]^T .\nonumber
\end{align}
Then, Eqs. (\ref{eq10})-(\ref{eq12}) can be written in compact form as
\begin{equation}
\label{eq16}
\dot{u}=A(t)u+\epsilon ,
\end{equation}
with the real time-dependent matrix
\begin{align}
\label{eq17}
A(t)=\left(
  \begin{array}{cccc}
    0 & \omega_m & 0 & 0\\
    -\Omega_m(t) & -\gamma & -G_x(t) & -G_y(t)\\
    G_y(t) & 0 & -\kappa & \Delta(t)\\
    -G_x(t) & 0 & -\Delta(t) & -\kappa
  \end{array}
\right),
\end{align}
where
\begin{equation}
\label{eq18}
\Delta(t)=\omega_c(q_0)-\omega_l
\end{equation}
is the effective detuning, $G_x(t)$ and $G_y(t)$ are, respectively, real and imaginary parts of the effective coupling
\begin{equation}
\label{eq19}
G(t)=\sqrt{2}q_{z}\omega_c^\prime(q_0)\alpha =G_x(t)+iG_y(t),
\end{equation}
and
\begin{equation}
\label{eq20}
\Omega_m(t)=\omega_m+q_{z}^2\omega_c^{\prime\prime}(q_0)|\alpha|^2.
\end{equation}
Since $\hat{\xi}$ and $\hat{a}_{in}$ are zero-mean quantum Gaussian noises and Eq. (\ref{eq16}) is linearised, the state of the quantum fluctuations converges to a time-dependent Gaussian state \cite{Gaussian-state}, fully characterized by the covariance matrix $V(t)$ whose elements are defined as
\begin{equation}
\label{eq21}
V_{ij}(t)=\left\langle u_i(t)u_j(t)+u_j(t)u_i(t)\right\rangle/2.
\end{equation}
From Eq. (\ref{eq16}), we can easily obtain an equation of evolution of $V(t)$:
\begin{equation}
\label{eq22}
\dot{V}(t)=A(t)V(t)+V(t)A^T(t)+D,
\end{equation}
where $D$ is the diffusion matrix of the noise and is defined as:
\begin{equation}
\label{eq23}
\delta(t-t^\prime)D_{ij}(t)=\left\langle \epsilon _i(t)\epsilon _j(t^\prime)+\epsilon _j(t^\prime)\epsilon _i(t)\right\rangle/2.
\end{equation}
\begin{figure*}[htp]
\centering
\includegraphics[width=6 in]{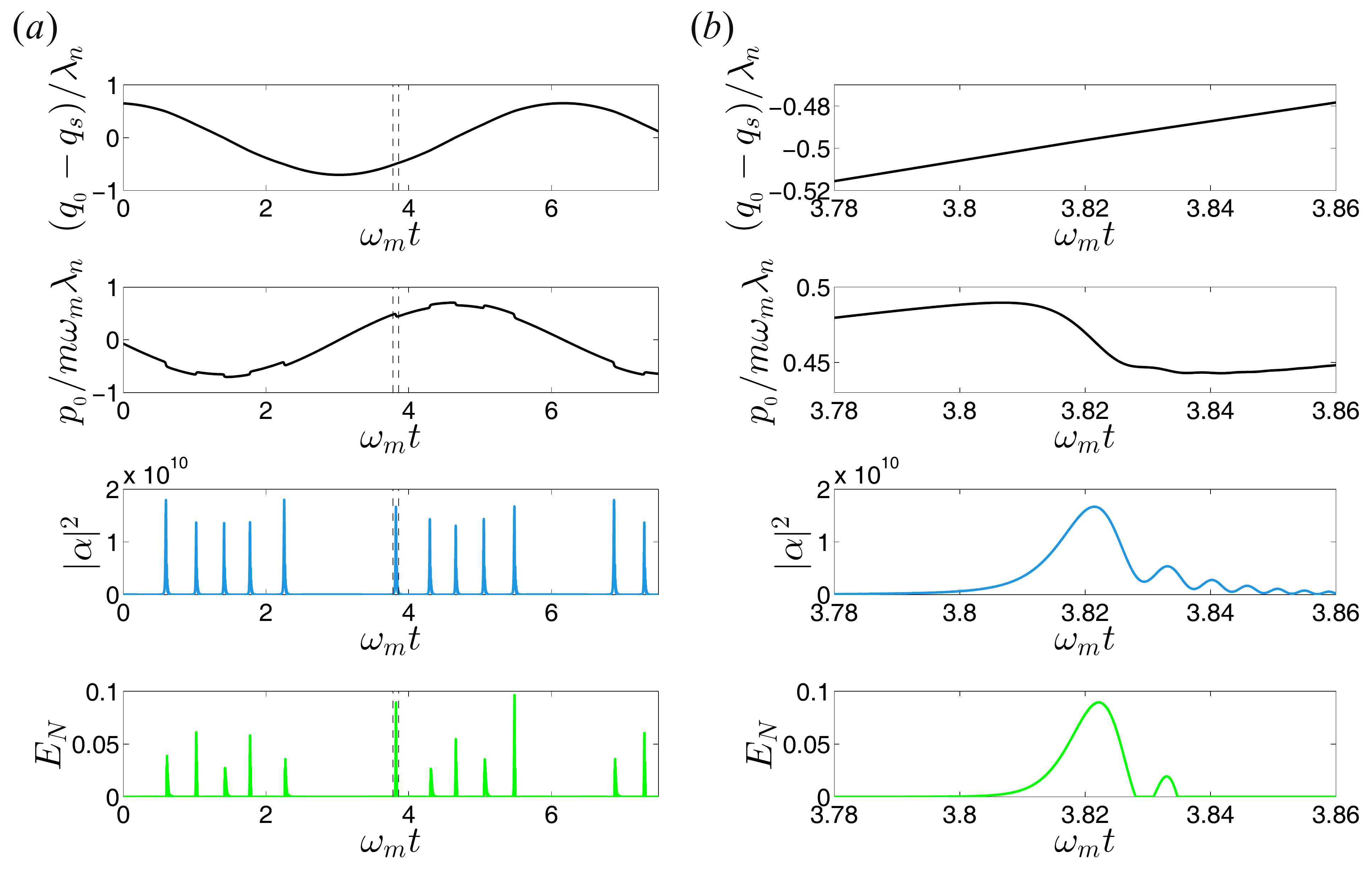}
\caption{(Color online) (a) Time evolution of $q_0$, $p_0$, $|\alpha|^2$ and $E_N$. (b) The respective zoom-in regions between the two vertical black dashed lines in (a). Here, the power of the driving laser is $P=0.21$ W, the membrane oscillates with the smallest limit cycle, and the temperature of the membrane is $T=1$ mK.
}
\label{fig7}
\end{figure*}
From Eq.(\ref{eq13}) and Eq.(\ref{eq15}), we can get that:
\begin{equation}
\label{eq24}
D=\text{diag}[0,\gamma(2n_{th}+1),\kappa,\kappa].
\end{equation}
The quantum entanglement between the optical cavity mode and the membrane can be quantified by the logarithmic negativity $E_N$, a quality which has been proved as a measure of entanglement \cite{entanglement-1}.
In the case of continuous variables, $E_n$ can be defined as \cite{entanglement-2}
\begin{equation}
\label{eq25}
E_N=\max[0,-\ln(2\eta^-)],
\end{equation}
where
\begin{align}
\eta^{-}&=2^{-1/2}\{\Sigma-[\Sigma ^2-4\det V]^{1/2}\}^{1/2},\nonumber\\
\Sigma &=\det V_1+\det V_2-2\det V_3,\nonumber
\end{align}
where $V_1$, $V_2$, and $V_3$ are $2\times2$ subblock matrices of $V(t)$ as
\begin{align}
\label{eq26}
V(t)=\left(
  \begin{array}{cc}
    V_1 & V_3\\
    V_3^T & V_2
  \end{array}
  \right).\nonumber
\end{align}
By numerically integrating Eqs. (\ref{eq7})-(\ref{eq9}) together with Eq. (\ref{eq22}), we can obtain the evolution of the quantum entanglement by using the logarithmic negativity.
We show the variation of the quantum entanglement with the mechanical oscillation of the membrane in Fig. \ref{fig6}.
Similar as the cavity occupation, the quantum entanglement also experiences a maximum value at positions $q_0=q_s+k\lambda_n/4$.

For more clearly compare the classical dynamical process and the evolution of the quantum entanglement, we plot the evolution of $q_0$, $p_0$, $|\alpha|^2$ and $E_N$ together in the time domain, as shown in Fig. \ref{fig7}.
It can be seen that, there is some synchronism between the classical dynamical process and the evolution of the quantum entanglement.
Every time the membrane passes though the positions $q_0=q_s+k\lambda_n/4$, the frequency of the cavity mode $\omega_c(q_0)$ meets the frequency of the external driving lasr $\omega_l$. The cavity mode is excited and the cavity occupation experiences a maximum value. The membrane sharply accelerates or decelerates due to the kick effect of the radiation pressure. The magnitude of the effective coupling $G(t)=\sqrt{2}q_{z}\omega_c^\prime(q_0)\alpha$ increases, and as a result, the quantum entanglement experiences a maximum value.
While, when the membrane moves away from the positions $q_0=q_s+k\lambda_n/4$, the cavity occupation decreases rapidly to almost zero. The membrane carries out damped harmonic oscillation. The magnitude of the effective coupling $G(t)$ decreases to almost zero, so the quantum entanglement vanishes due to the decoherence.

%%%%%%%%%%%%%%%%%%%%%%%%%%%%%%%%%%%%%%%%%%%%%%%%%%%%%%%%%%%%%%%%%%%%%%%%%%%%%%%%%%%%%%
\section{Summary}
\label{sec5}

We have studied classical dynamics and quantum entanglement of a MIMOS in the ELAR, in which the optical cavity mode can be excited multiple times during one cycle of the mechanical oscillation of the membrane.
Every time the optical cavity mode is excited, the radiation pressure kicks the membrane and makes it sharply accelerate or decelerate.
So the membrane can present self-sustained oscillations with limit cycles in the shape of sawtooth-edged ellipses.
We have demonstrated that the membrane can exhibit dynamical multistability in a wide range of parametric space.
We have also studied the dynamics of the quantum fluctuations around the classical orbit and calculated the evolution of the quantum entanglement between the membrane and the optical cavity mode during the mechanical oscillation.
We have shown that there is some synchronism between the classical dynamical process and the evolution of the quantum entanglement, and this may find some applications in quantum information processing.

\no

% \newpage

% \newpage

% \noindent {\footnotesize{\bf Table 1}\quad Some typical quantities
% calculated in different models with several particle numbers $N$,
% where $H_{\rm cri}$ denotes the
% Hamiltonian at the critical point of the U(5)-O(6) transition}\vspace{-5mm}\\

\vspace*{2mm} \Acknowledgements{\bahao This work is supported by the National Natural Science Foundation of China under Grant Nos. 11175094 and 91221205 and the National Basic Research Program of China under Grant No. 2011CB9216002.}

\end{multicols}

\end{document}